\documentclass[fleqn,12pt,twoside]{article}
\usepackage{espcrc1}
\usepackage{graphicx}
\usepackage[figuresright]{rotating}

\newcommand{\AmS}{{\protect\the\textfont2
  A\kern-.1667em\lower.5ex\hbox{M}\kern-.125emS}}
\title{Experimental Review of Baryons in the Nuclear Medium}
\author{
    S.~Schadmand
    \address{
    Institut fü\"ur Kernphysik,
    Forschungszentrum J\"ulich,
    D-52425 J\"ulich, Germany
    }
    \thanks{
    Work supported by
    Deutsche Forschungsgemeinschaft,
    UK Science and Engineering Research Council, and
    Schweizerischer Nationalfonds.
    }
}

\begin{document}
\maketitle

\begin{abstract}
Inclusive studies of nuclear photoabsorption have provided
clear evidence of medium modifications but the results have not yet been
explained in a model independent way.
A deeper understanding of the situation
is anticipated from a detailed experimental study of meson photoproduction from
nuclei in exclusive reactions.
Recent results on meson production in photonuclear experiments
indicate a large difference between quasi\-free meson
production from the nuclear surface and non-quasifree components.
\end{abstract}

\section{INTRODUCTION}

Photoabsorption experiments on the free nucleon  demonstrate the complex structure
of the nucleon and its excitation spectrum.
\begin{figure}[htb]
\begin{minipage}{0.5\textwidth}
\includegraphics[width=\textwidth]{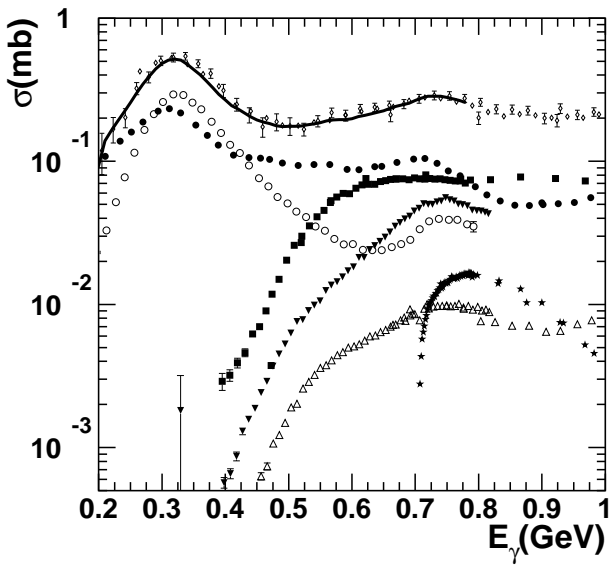}
\end{minipage}
\hfill
\begin{minipage}{0.45\textwidth}
\vspace*{-8mm}
\caption{
 Photoabsorption cross section on the proton
 and decomposition into meson production channels.
 Small open circles are the photoabsorption data compilation from~\cite{Hagiwara:2002fs}.
 The experimental meson production cross sections are:
 single $\pi^+$     (solid circles) from~\cite{Buechler:1994jg,MacCormick:1996jz},
 single $\pi^\circ$ (open circles) from~\cite{Harter:1997jq},
 $\pi^+\pi^-$ (solid squares) from~\cite{Braghieri:1995rf,Wisskirchen-doc},
 $\pi^+\pi^\circ$  (downward solid triangles) from~\cite{Langgartner:2001sg},
 $\pi^\circ\pi^\circ$ (upward open triangles) from~\cite{Kotulla-doc,Hourany-inpc-2001},
 and
 $\eta$ (stars) from~\cite{Krusche:1995nv,Renard:2000iv}.
 The solid line is the sum of the meson channels up to 800~MeV.
}\label{fig:photo-decomp-p}
\end{minipage}
\end{figure}
The lowest resonance is called $\Delta$(1232) which is a P$_{33}$ state
in the common notation (L$_{(2I)(2J)}$) with a pole mass of 1232~MeV.
It is prominently excited by incident photons of 0.2--0.5~GeV.
The following group of resonances, P$_{11}$(1440), D$_{13}$(1520), and S$_{11}$(1535),
is called the second resonance region (E$_\gamma$=0.5-0.9~GeV).
The observed resonance structures
have been studied using their decay via light mesons, showing that the photoabsorption
spectrum can be explained by the sum of $\pi$, $\pi\pi$ and
$\eta$ production cross sections.
Fig.~\ref{fig:photo-decomp-p}
shows the photoabsorption cross section
on the proton along with the experimental meson photoproduction cross sections.
The shapes of the meson cross sections reflect the resonance structures
observed in photoabsorption showing that the mesons are mostly decay products
of the respective resonances.
Single pion production is dominant in the region of the $\Delta$(1232) resonance.
Also, the three resonances
comprising the second resonance region, decay to $\sim$50\% via single pion emission.
This fact has been extensively exploited in partial wave analyses.
Above E$_\gamma\approx$0.4~GeV, the photoproduction of two pions is kinematically
possible and single $\pi$ production looses in dominance.
The solid line in Fig.~\ref{fig:photo-decomp-p} represents
the sum of the meson cross sections up to 0.8~GeV and demonstrates
that the photoabsorption cross section on the proton can be
explained by its decomposition into meson production.

The $\eta$ production threshold is located at E$_\gamma\approx$700~MeV.
The steeply rising $\eta$ cross section in Fig.~\ref{fig:photo-decomp-p}
is characteristic for an s-wave resonance.
The angular distributions of the $\eta$ emission are consistent
with this observation and
the cross section peaks around the mass pole of the S$_{11}$(1535)
resonance~\cite{Krusche:1995nv,Renard:2000iv}.
This resonance is unique in the sense that is has a strong decay branch
of 30-55\% into $\eta$ mesons.
Thus, $\eta$ production is considered characteristic for the S$_{11}$(1535) resonance.
Above the $\eta$ threshold, the cross section basically displays the resonance line shape
enabling detailed studies of that state.
However, the contribution to the total is small.

\begin{figure}[htb]
\centering
\includegraphics[width=0.45\textwidth]{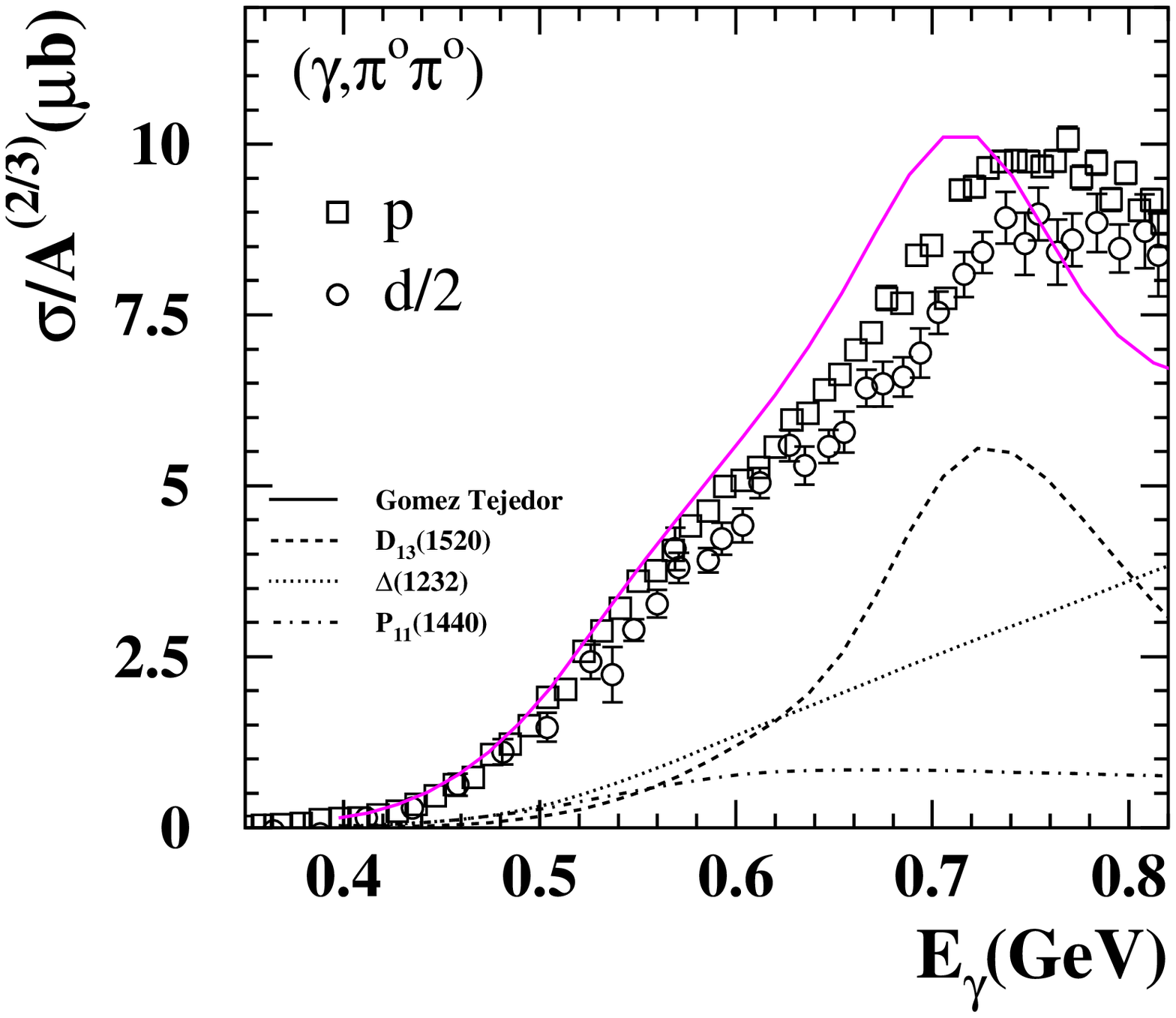}
\includegraphics[width=0.45\textwidth]{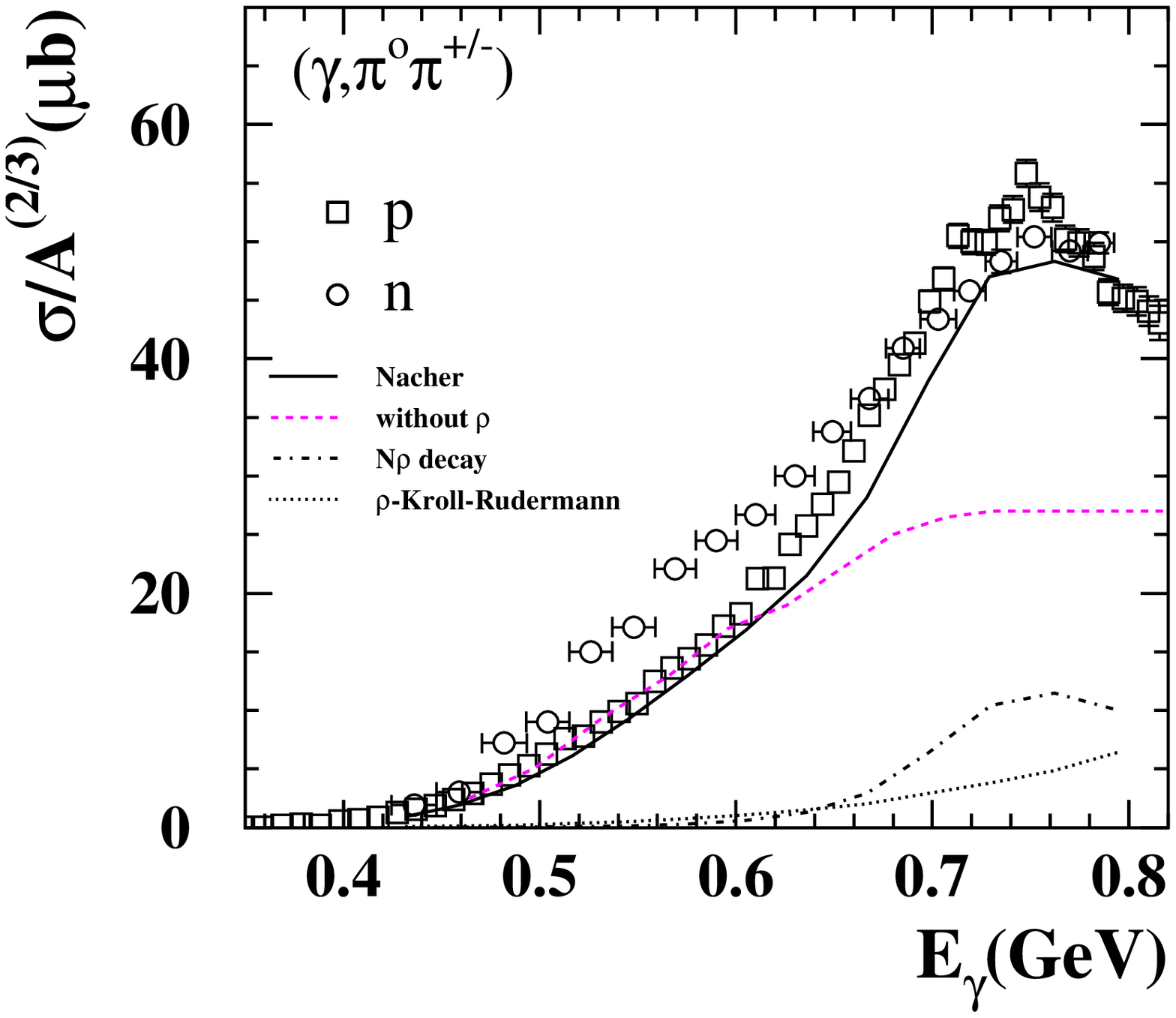}
\caption{
Double pion production from the nucleon.
The curves are calculations from \cite{GomezTejedor:1996pe,Nacher:2000eq}.
Left: comparison of
2$\pi^\circ$ from the proton and deuteron \cite{Krusche:2003ik}.
Right: comparison of
$n\pi^\circ\pi^+$ \cite{Braghieri:1995rf,Langgartner:2001sg} with $p\pi^\circ\pi^-$
\cite{Zabrodin:1997xd}.
}
\label{fig:pipi-nucleon}
\end{figure}
In the second resonance region,
the dominant contribution comes from the D$_{13}$(1520) resonance
which has the strongest coupling to the incident photon.
Single as well as double pion production channels display
structure at the corresponding resonance mass,
i.e. around E$_\gamma\approx$760~MeV.
Fig~\ref{fig:pipi-nucleon} shows results on the photoproduction
of pion pairs on the nucleon.
As could already be seen from Fig.~\ref{fig:photo-decomp-p},
the meson production channels involving charged pions are dominant
as expected in electromagnetic excitation processes.
On the proton, three isospin combinations of pion pairs
can be produced.
$\pi^\circ\pi^\circ$ and $\pi^+\pi^-$ production revealed
that the resonance decays sequentially via an intermediate
$\Delta$ state~\cite{Braghieri:1995rf,Wolf:2000qt}.
In $\pi^+\pi^\circ$ the same behavior was found.
In addition, a decay branch of 20\% $N^\star\to N\rho$
was deduced~\cite{Langgartner:2001sg,Nacher:2000eq}.
On account of its dominance, the D$_{13}$(1520) resonance
is said to be tagged by double pion production.
However, calculations by Gomez Tejedor and Oset~\cite{GomezTejedor:1996pe},
showed that the $N^*$ contribution to double pion photoproduction
by itself is not large but rather stems from an interference with
other terms.

\section{NUCLEAR PHOTOABSORPTION}

\begin{figure}[hbt]
    \includegraphics[width=0.45\textwidth]{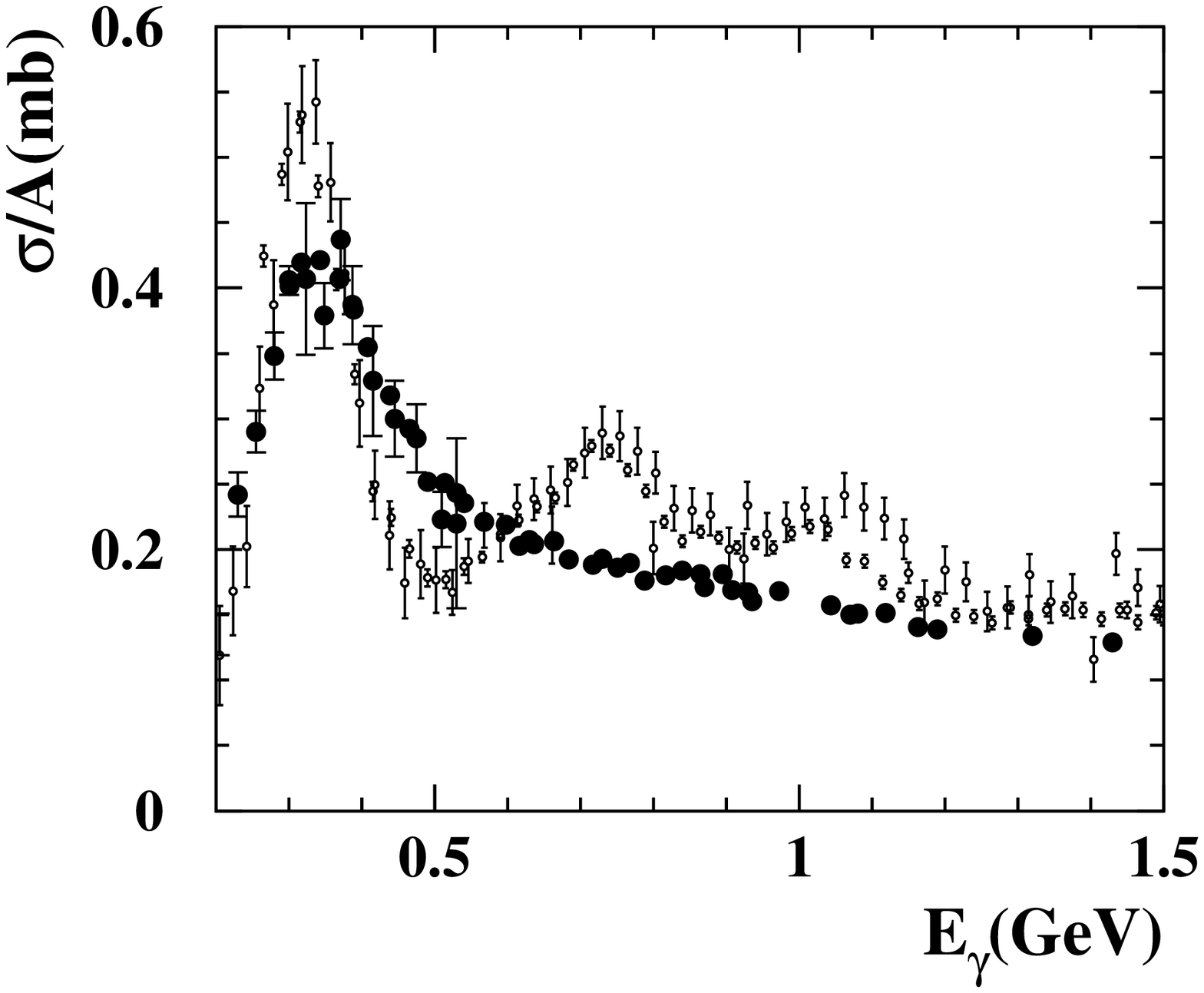}
\hfill
    \includegraphics[width=0.5\textwidth]{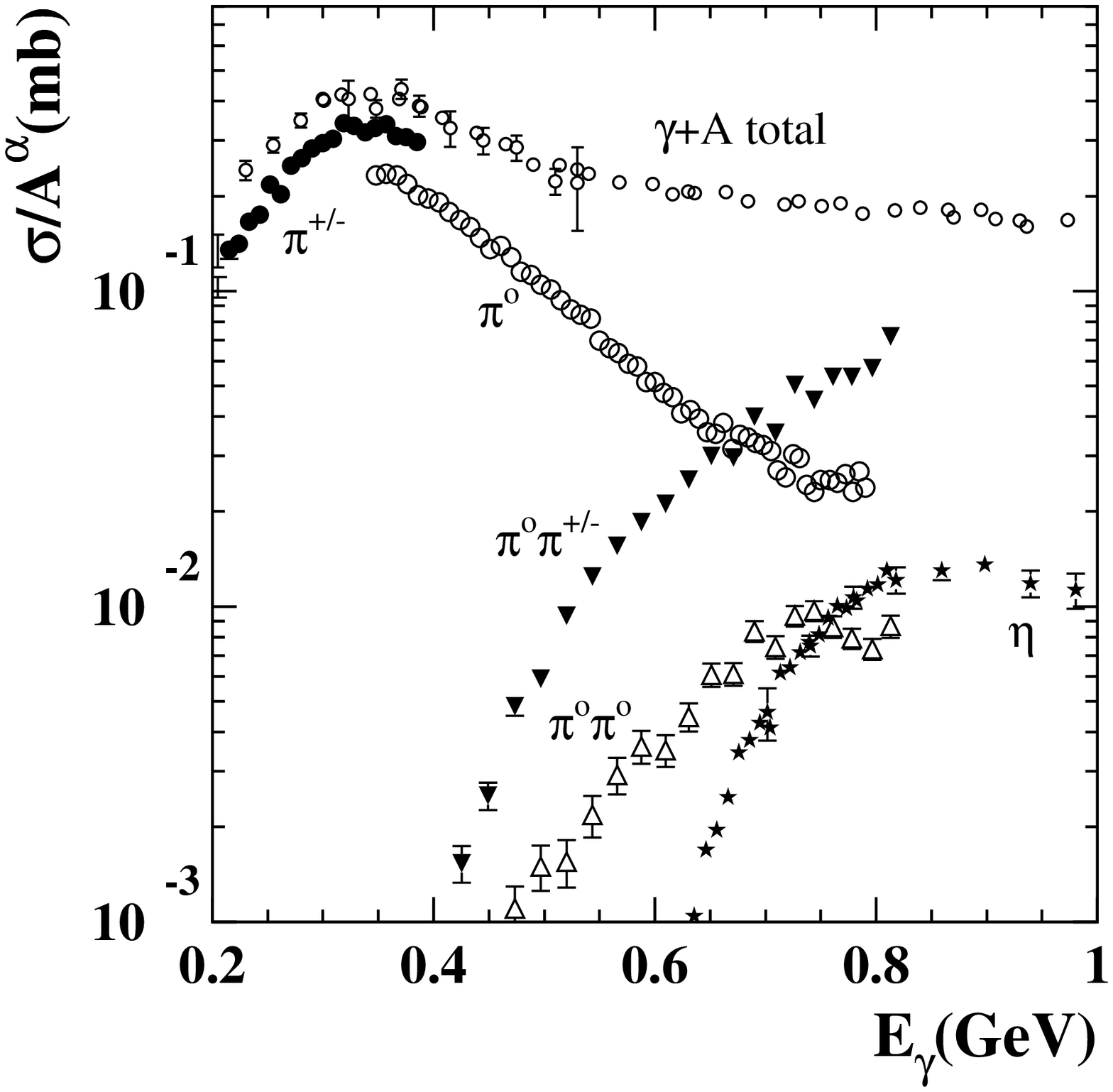}
\caption{
Left:
Nuclear photoabsorption cross section per nucleon
as an average over the nuclear systematics~\cite{Muccifora:1998ct} (full symbols)
compared to the absorption on the proton~\protect\cite{Hagiwara:2002fs} (open symbols).
Right:
  Status of the decomposition of nuclear photoabsorption into
  meson production channels (scaled with A$^\alpha$, $\alpha$=2/3).
  Small open circles are the average nuclear photoabsorption cross section
  per nucleon ($\alpha$=1) \protect\cite{Muccifora:1998ct}.
  Meson production data are from
  \protect\cite{Arends:1982ed,Krusche:2001ku,Roebig-Landau:1996xa,Yamazaki:2000jz,Janssen_02}.
}\label{fig:photoabs-nucs}
\end{figure}

The left panel of
Fig.~\ref{fig:photoabs-nucs} shows the nuclear photoabsorption cross section per nucleon
as an average over the nuclear systematics~\cite{Muccifora:1998ct}.
\newline
The $\Delta$ resonance is broadened and slightly shifted while the second and higher
resonance regions seem to have disappeared.
The right panel shows the status of the decomposition of nuclear photoabsorption into
meson production channels. The available experimental meson cross sections are exclusive
measurements, investigating quasifree production. It can be inferred from the figure that
the sum of the cross sections, which are missing the purely charged final states, cannot
reproduce the shape of the total photoabsorption from nuclei.

\begin{figure}[htb]
 \includegraphics[width=0.45\textwidth]{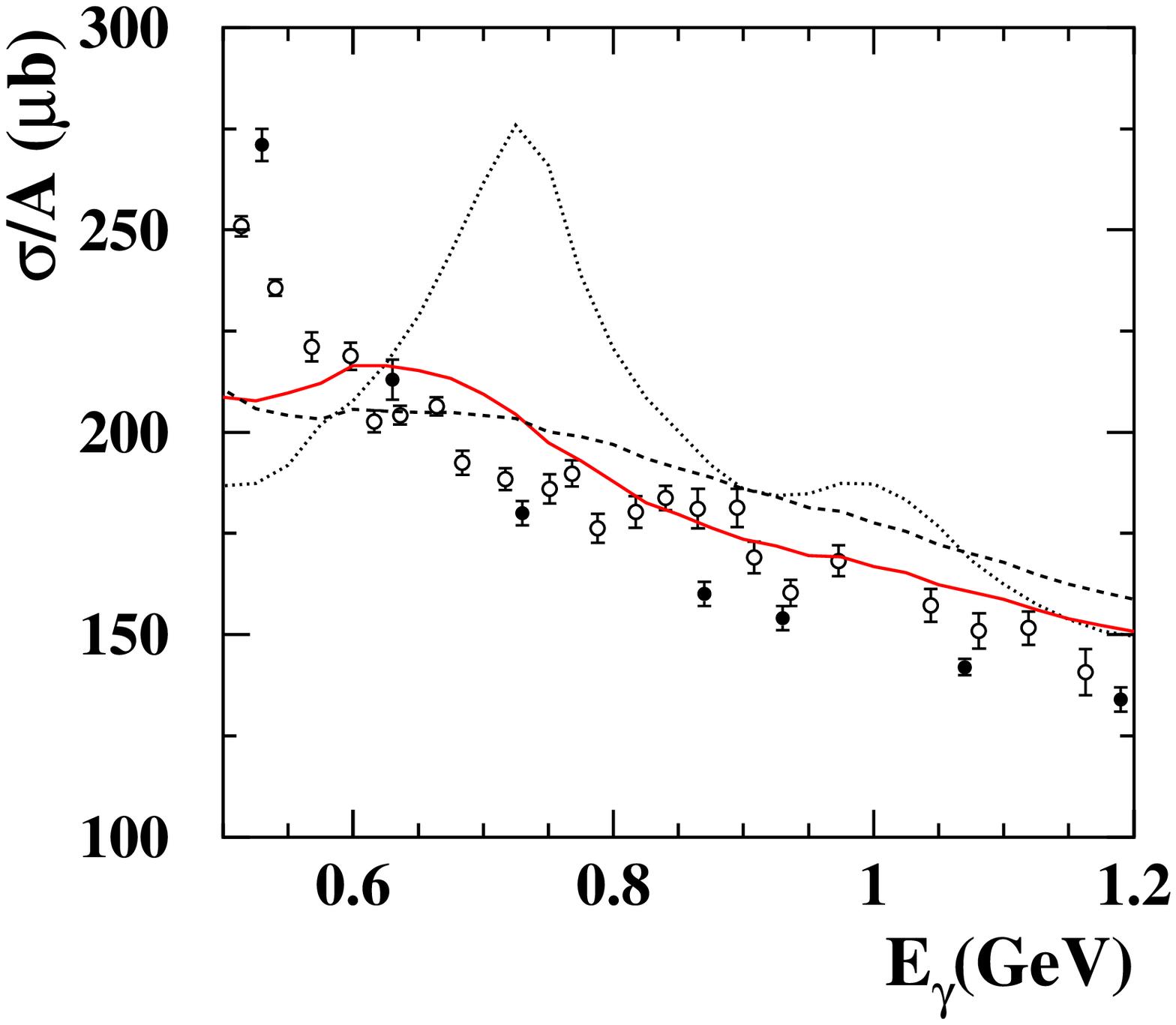}
 \hfill
 \includegraphics[width=0.4\textwidth]{photoabs-hirata.eps}
 \caption{
 Left:
    Modification of the second resonance region for the
    $\gamma^*{}^{40}\textrm{Ca}\to X$-cross section from BUU calculations \cite{Lehr:1999zr}.
    The solid curve is calculated with an explicit enhancement of the $N\rho$ width,
    the dashed curve with a momentum dependent potential.
    Dotted curve: photoabsorption on the free proton.
    Experimental data are from \cite{Bianchi:1999gs}.
Right:
     Total nuclear photoabsorption cross section on nuclei according to Hirata et al.
    \cite{Hirata:2001sw}.
    The solid curve is the full calculation.
    Individual contributions of the absorption processes are shown by themselves:
    one pion production (dash-dotted line),
    two pion production (dashed line),
    many body absorption through the $\Delta$-nucleus state and $N^*$-nucleus state (dotted line), and
    many body absorption through the $\pi\Delta$-nucleus state (long dashed line).
 }\label{fig:photoabs-theo}
\end{figure}

Mosel et al.~\cite{Mosel:1998rh},
have argued that an in-medium broadening of the D$_{13}$(1520)
resonance is a likely cause of the suppressed photoabsorption cross
section.
The calculation is based on the BUU equation which describes the
space time evolution of the spectral phase space density
of an ensemble of interacting particles.
For the baryons, the mean field potential is determined as described
in \cite{Lehr:1999zr}. Here, nuclear incompressibilities corresponding to
a hard equation of state (momentum independent potential) and
a medium equation of state (momentum dependent potential) are employed.
The photon-nucleus reaction is modelled in terms of the absorption
of the photon on a single nucleon (quasifree process).
Over a more comprehensive energy range, this leads to final states P$_{33}$(1232),
D$_{13}$(1520), S$_{11}$(1535), F$_{15}$(1680), $N\pi$, $N\pi\pi$,
$NV$, $K\Lambda$, $K\Sigma$ and $K\bar K N$.
The states are prepared according to the respective cross sections.
Final state interactions are described by a set of BUU equations.
The cross sections for the $\gamma A$ reaction are determined by averaging
over an ensemble of such elementary reactions as outlined in
\cite{Effenberger:1997rc}.
Besides Fermi motion, binding effects and Pauli blocking,
collisional broadening of the most important resonances P$_{33}$(1232),
D$_{13}$(1535) and S$_{11}$(1535) is accounted for.
The left panel of Fig.~\ref{fig:photoabs-theo} shows the total photonuclear
cross sections where different scenarios are presented.
The use of a momentum dependent $N^*$ potential leads to a
smearing of the D$_{13}(1520)$ because the momentum dependence
leads to a shift of the effective mass.
For a D$_{13}(1520)$ produced with momenta around 800 MeV, the
nucleon potential almost vanishes.
This shift amounts to $\Delta m^*\approx 50$ MeV.
The broadening is due to the strong increase of the width with the mass.
An almost complete disappearance of the resonant structure is observed
but the experimental data are still overestimated.
Altering the resonant contribution by using an enhanced
in-medium width leads to a more smeared and reduced cross section.
An enhancement of the $N\rho$ width at nuclear matter density by about a factor
10 was found with a total width at the pole mass of about 335 MeV.
The use of this in-medium $N\rho$ width includes the full mass, momentum,
and density dependence and leads to the solid curve in Fig.~\ref{fig:photoabs-theo}.
The description of the experimental data is considerably improved.
However, a bump structure survives for photon energies around 650 MeV
which is caused by the strong mass dependence of the
D$_{13}$ width which is reflected in the in-medium width.

Hirata et al.~\cite{Hirata:2001sw}
have argued that a change of the interference effects in the
nuclear medium is one of the most important reasons for the suppression of
the resonance structure.
The right panel of Fig.~\ref{fig:photoabs-theo}
shows the calculated cross section per nucleon.
The contributions of one pion and two pion production, many body absorption
through the $\Delta$-nucleus and $N^*$-nucleus state, as well as
many-body absorption through the $\pi\Delta$-nucleus state are
shown by themselves.
Here, two-pion photoproduction is about 3 times smaller
than in the elementary process due to cooperative effects between
the different medium effects as in
spreading potentials for $\Delta$ and $N^*$, pion distortion,
and modified interferences among the related reaction processes.
The cross sections of the other many-body processes are almost
flat in the energy range above 600 MeV and small.
The excitation peak around the position of the $N^*$ resonance
in the total nuclear photoabsorption cross section is indeed not
present in the calculation.
However, the model underestimates the nuclear cross section in the
valley region between 380 and 500 MeV by about 15 percent.
It is inferred that there must be an important processes enhancing nuclear
photoabsorption in the valley region.
Here, the intermediate pion and $\rho$ meson are far off-shell and
two nucleons could explicitly contribute.
In addition, the cross sections are underestimated slightly at the $\Delta$
resonance energy around 320 MeV as
coherent $\pi^0$ production is not included in the calculation.

It may be concluded that inclusive reactions like total photoabsorption
do not allow a detailed investigation of in-medium effects.
A deeper understanding of the situation is anticipated from the
experimental study of meson photoproduction on
nucleons embedded in nuclei in comparison to studies on the free nucleon.

\section{MESON PHOTOPRODUCTION FROM NUCLEI}

In reactions from the free nucleon, baryon resonance properties are
extracted by tagging on their characteristic meson decay.
The same procedure may be applied to nucleon excitations in the nuclear
medium.
In the second resonance region, double pion production aims at
the resonances D$_{13}$(1520) and P$_{11}$(1440) while
$\eta$ production is characteristic for the S$_{11}$(1535) resonance.
As pointed out above, the three resonances in the second resonance
region decay to roughly 50\% via single pion emission.

The most trivial medium modification is the broadening of the excitation
functions due to Fermi motion.
The decay of the resonances is further modified by Pauli-blocking
of final states, which reduces the resonance widths.
In addition, decay channels like $\mbox{N}^{\star}\mbox{N}\rightarrow \mbox{NN}$
cause collisional broadening.
Both effects cancel to some extent and it is a priori not clear
which will dominate.

\begin{figure}[hbt]
\includegraphics[width=0.45\textwidth]{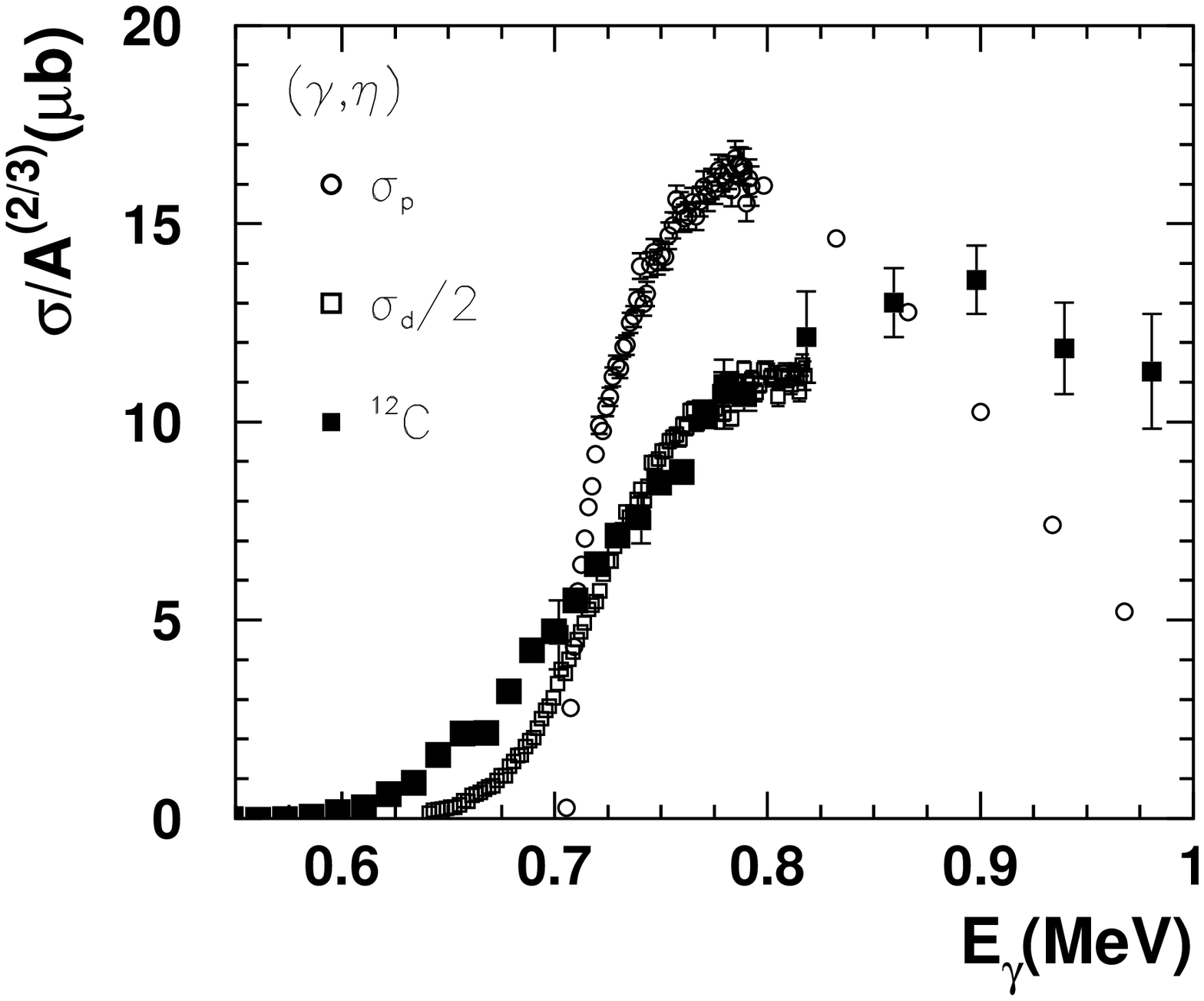}
\hfill
\includegraphics[width=0.5\textwidth]{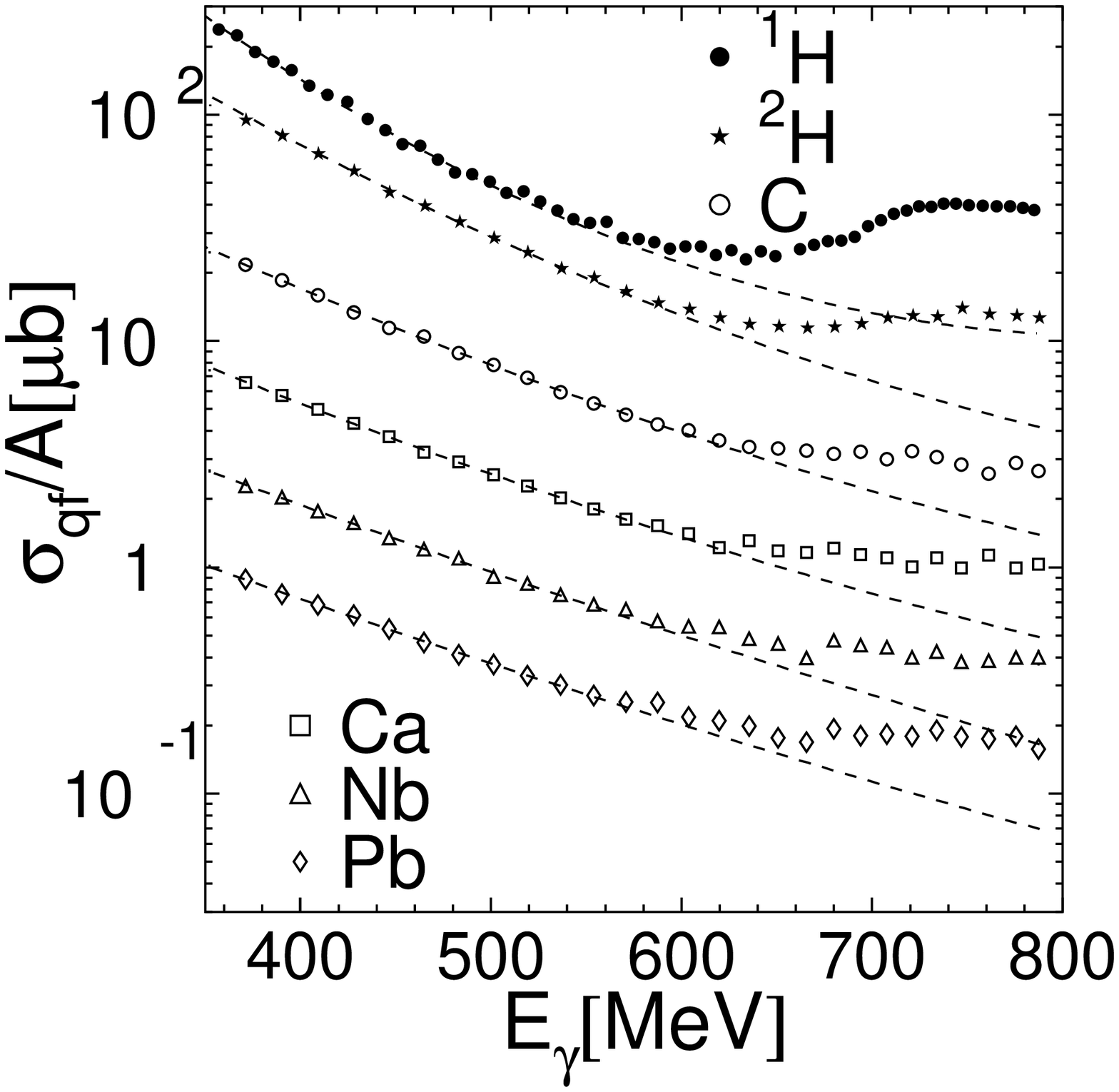}
\caption{
Left:
Comparison the total cross section
per nucleon for $\eta$ photoproduction on nucleons \cite{Krusche:1995nv,Renard:2000iv}
and using a carbon target \cite{Roebig-Landau:1996xa,Yamazaki:2000jz}.
Right:
Total cross section per nucleon for single $\pi^\circ$
photoproduction in the second resonance region for the nucleon and for nuclei.
The scale corresponds to the proton data, the other data are scaled down
by factors 2,4,8,16,32, respectively.
The dashed curves are fits to the data in the energy range 350--550~MeV.
}\label{fig:bernd-nucs}
\end{figure}

\subsection{$\eta$ PRODUCTION}

On the free proton, the photoproduction of $\eta$-mesons in the second resonance region
proceeds almost entirely through the excitation of the S$_{11}$(1535) resonance.
An observation of the reaction over a series of
nuclei~\cite{Roebig-Landau:1996xa,Yamazaki:2000jz} did not show
a depletion of the in-medium strength.
The left panel of Fig.~\ref{fig:bernd-nucs} compares the total cross section
per nucleon for $\eta$ photoproduction on the proton and using a carbon target.
This result is in line with theoretical findings
that the change of the S$_{11}$ self energy in the medium is
small~\cite{Post:2003hu,Inoue:2002xw}
and are in agreement with model calculations that take the
trivial in-medium effects and final state interactions
into account \cite{Effenberger:1997rc,Carrasco:1993sk}.
A recent study  shows that the data
could be described over the full energy range by applying a momentum
dependent S$_{11}$ potential \cite{Lehr:2003km}.

\subsection{SINGLE $\pi^\circ$ PRODUCTION}

An attempt to study the in-medium properties of
the D$_{13}$ resonance was undertaken with a measurement of quasifree single
$\pi^\circ$ photoproduction \cite{Krusche:2001ku} which, on the free nucleon,
is almost exclusively sensitive to the D$_{13}$ resonance.
The right panel of Fig.~\ref{fig:bernd-nucs} summarizes the results.
Strong quenching of the D$_{13}$-resonance structure
is found for the deuteron with respect to the nculeon.
However, an indication of a broadening or a suppression
of the D$_{13}$ structure in heavy nuclei is not observed.
Model predictions agree with the pion photoproduction
data only under the assumption of a strong broadening of the resonance,
other effects seem to be missing in the models. This casts doubt
on the interpretation of the total photoabsorption data via resonance
broadening.
In contrast to the case of total photoabsorption, the second resonance bump remains visible.
However, exclusive reaction channels are dominated by the nuclear surface region
where in-medium effects are smaller.
Furthermore, as discussed in
\cite{Lehr:2001ju}, resonance broadening effects are even more diluted for reactions
which do not contribute to the broadening, due to the averaging over the nuclear volume.

\subsection{DOUBLE PION PRODUCTION}

\begin{figure}[ht]
 \centering
  \includegraphics[width=0.48\textwidth]{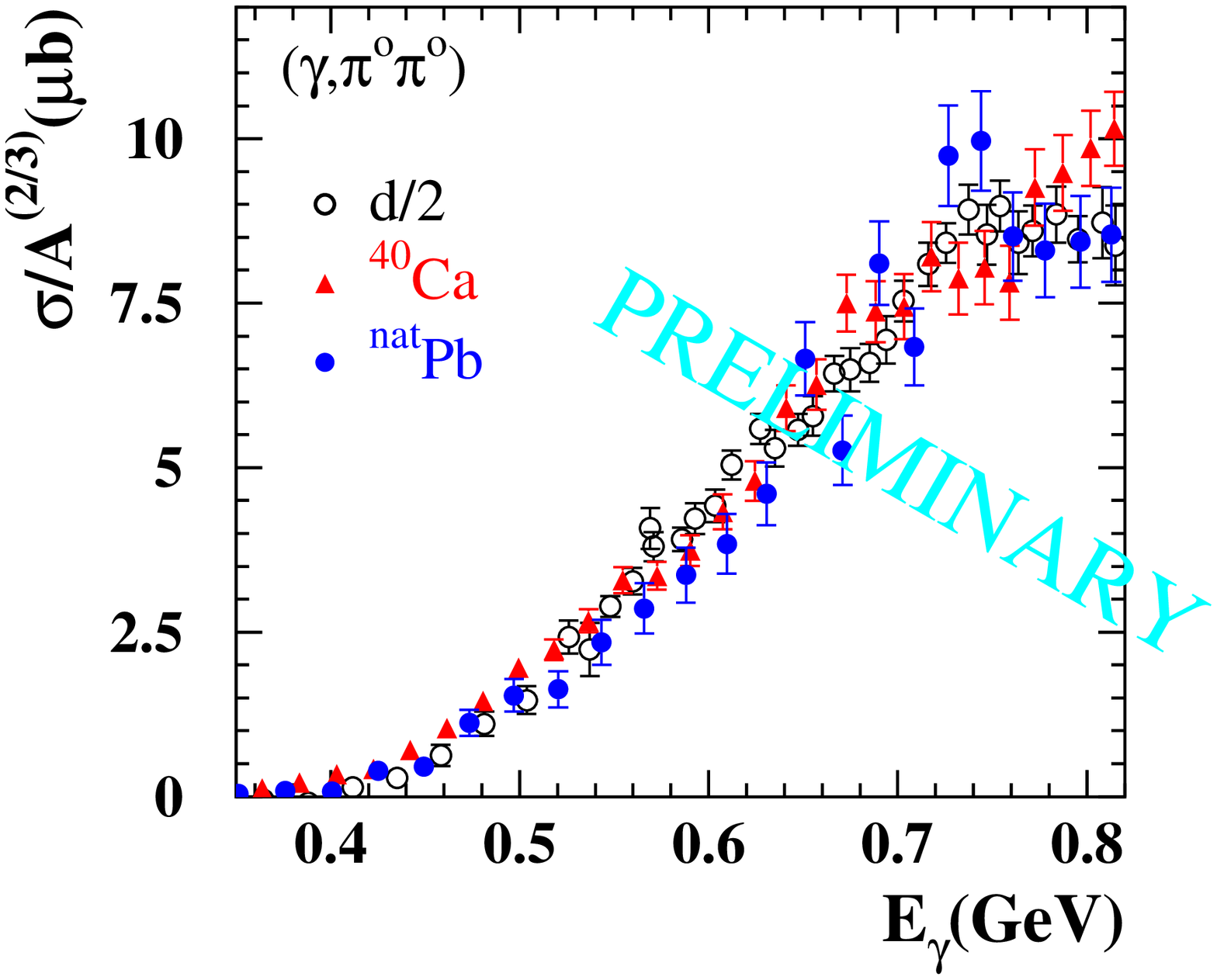}
  \includegraphics[width=0.48\textwidth]{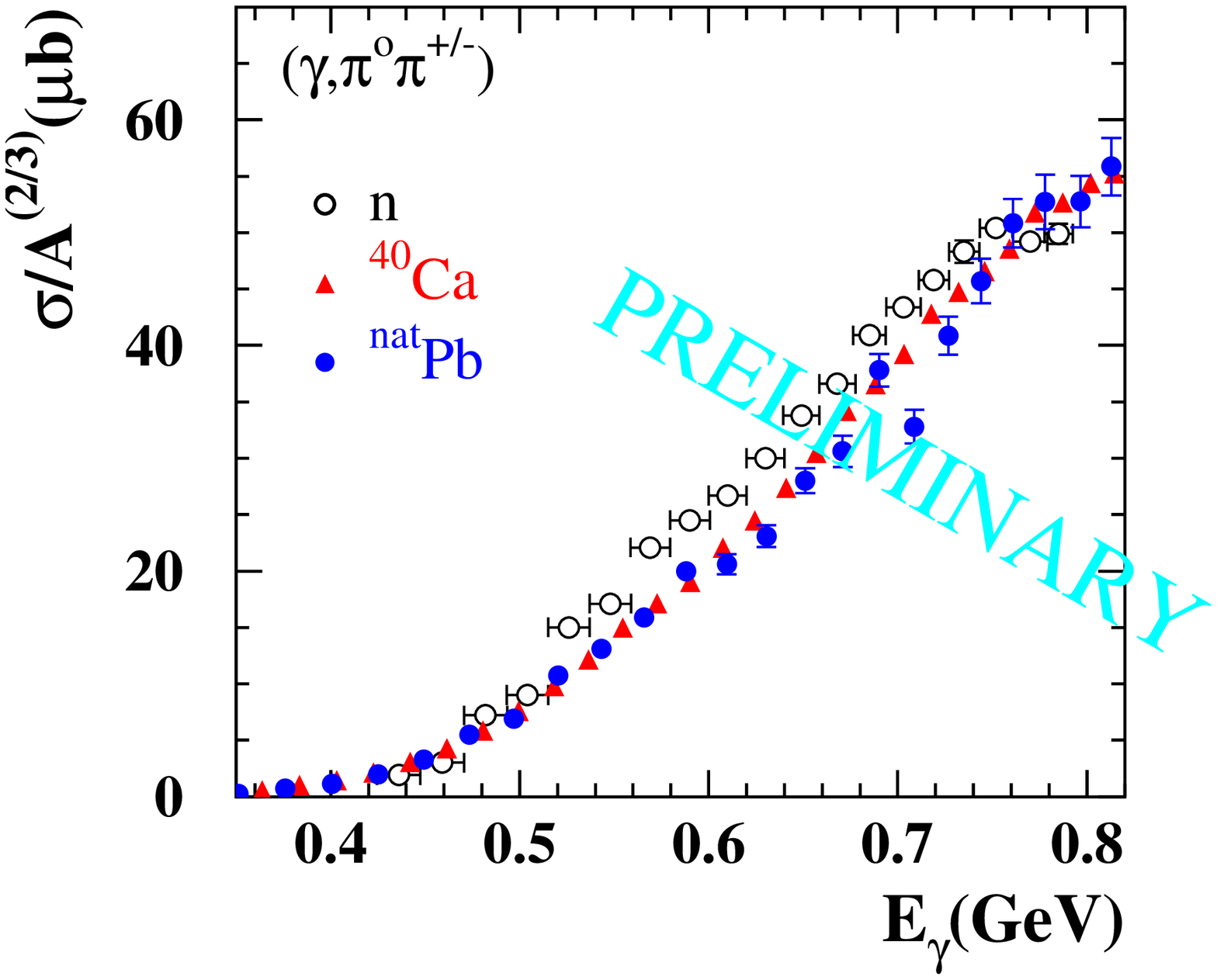}
 \caption{
    Preliminary total cross sections for $\pi\pi$
    photoproduction from lead~\protect\cite{Janssen_02} along with
    results from the deuteron~\protect\cite{Zabrodin:1997xd,Kleber:2000qs}.
    The nuclear cross sections are divided by A$^{2/3}$,
    the $\pi^\circ\pi^\circ$ deuteron cross section by 2.
}\label{fig:pipi-nucs}
\end{figure}

Fig.~\ref{fig:pipi-nucs} shows preliminary cross sections for
$\pi^\circ\pi^\circ$ and $\pi^\circ\pi^\pm$ photoproduction
on calcium and lead from a recent TAPS analysis.
The nuclear cross sections are divided by A$^{2/3}$ and
compared to results from the free proton and from nucleons bound in deuterons.
With the scaling with A$^{2/3}$, the nuclear data agree almost exactly
with the cross sections on the nucleon.
Thus, the total nuclear $\pi\pi$ cross sections
do not seem to show any modification beyond absorption effects.
It may be speculated that the strong 2$\pi$ decay branch
via $\Delta$ intermediate states ($N^\star\to\Delta\pi\to N\pi\pi$),
together with the fact that the $\Delta$ resonance itself does not
dramatically change in medium, dominate this behavior.
Also, in the reaction $\pi^\circ\pi^\pm$,
the two pions can stem from the decay of the $\rho$ meson while
the decay $\rho\to\pi^\circ\pi^\circ$ is forbidden.
Accordingly, detailed studies of differential cross sections might reveal
different modifications of the $\pi\pi$ correlations.
A first result came from the investigation of $\pi\pi$ invariant mass
distributions in the incident photon energy range of
400--460~MeV~\cite{Messchendorp:2002au} providing indication of
an effect consistent with a significant in-medium
modification in the $A(\gamma,\pi^\circ\pi^\circ)$ ($I$=$J$=0) channel.

\subsection{SUMMARY}

\begin{figure}[htb]
\centering
\includegraphics[width=0.6\textwidth]{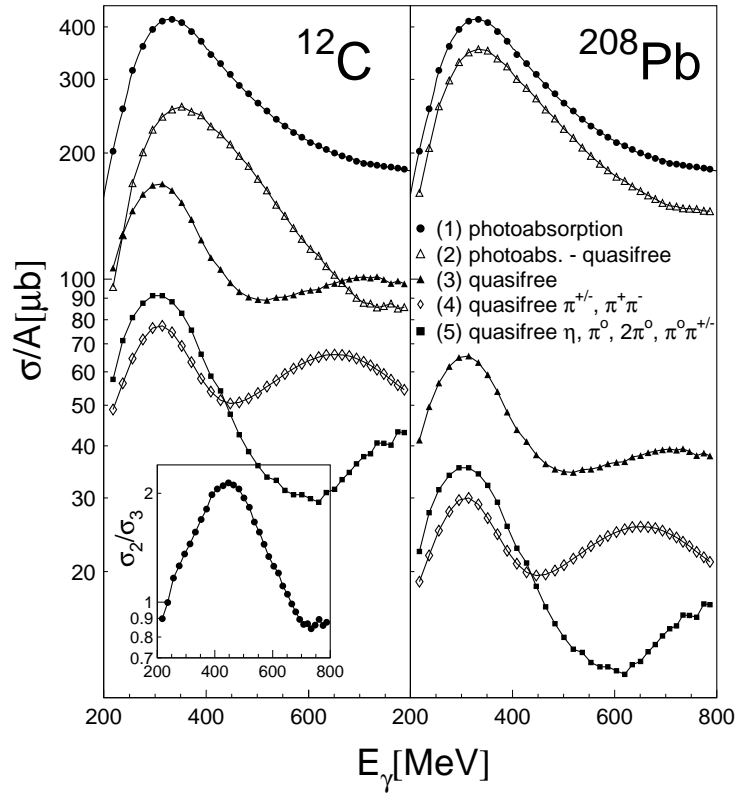}
\caption{
Decomposition of the total photoabsorption cross section (1) for carbon (left)
and lead (right) \cite{Krusche:2004zc}.
The sum of quasifree cross sections (3) is built from an approximation of purely charged
final meson states (4) and the available cross sections containing
at least one neutral meson (5).
Curve (2) represents the difference of the estimated total quasifree meson production
from the total absorption cross section.
Inset: ratio of (2) and (3) for carbon.
}\label{fig:splitup}
\end{figure}

The information about meson photoproduction from nuclei was obtained
with the photon spectrometer TAPS in combination with the Glasgow tagged photon facility
at MAMI-B in Mainz.
Technically, the experimental setup was tuned towards the detection of neutral
mesons and charged pions could only be reliably detected in the presence
of at least one photon (thus, in the presence of a $\pi^\circ$),
from relative time-of-flight information.
Data on the photoproduction of purely charged meson final states from nuclei
are not available for the second resonance region.
However, the neutral quasifree reactions consistently follow the scaling
with the nuclear surface and charged pions will undergo similar final state effects.
Thus, it is reasonable to assume the same scaling behavior
and the quasifree cross section for charged meson production
from the deuteron cross section has been approximated
in Ref.~\cite{Krusche:2004zc}.
The deuteron cross section was folded with a typical momentum distribution
for nuclei in order to account for the stronger Fermi motion effects.

The result for carbon is shown in Fig.~\ref{fig:splitup}
(left side, curve (4)) together with the quasifree cross section for neutral
and mixed charged states (curve (5)), and the total photoabsorption
cross section  (curve (1)). The behavior for heavier nuclei
is qualitatively the same (see Fig.~\ref{fig:splitup}, right side).
Here, the all cross sections are scaled by $A$
and the inherent $A^{2/3}$ scaling of the quasifree reactions make the latter
relatively less important for heavier nuclei.
The sum of the quasifree meson production cross sections (curve (3)) shows
clear signs of the $\Delta$ resonance and the second resonance region.
The flattening is mainly due to Fermi motion effects.
The excitation function reflects the typical response of the
low density nuclear surface regions to photons.
The difference between this cross section and total photoabsorption represents the typical
response of the nuclear volume (curve(2)) where isolated resonance peaks
are not seen.
The inset in the figure shows the ratio of these two excitation functions
for carbon.
The most striking feature is the buildup of strength at incident photon
energies around 400 MeV in the volume component as compared to the quasifree
surface reactions. It is known  that two-body absorption
mechanisms like $\gamma$NN$\rightarrow N\Delta$ are non-negligible in
this energy range \cite{Carrasco:1992vq}, but it is not known if they can explain the effect.
Further progress in the models is necessary for an understanding of this
behavior.
Also, it would be desirable to complete the experimental picture by investigating
single charged pion as well as $\pi^+\pi^-$ production from nuclei.

\section{CONCLUSIONS}

The systematic study of total cross sections for single $\pi^\circ$,
$\eta$, and $\pi\pi$ production over a series of nuclei has not
provided an obvious hint for a depletion of resonance yield.
The observed reduction and change of shape in the second resonance region
are mostly as expected from absorption effects, Fermi smearing and Pauli blocking,
and collisional broadening.
The sum of experimental meson cross sections
for neutral and mixed charged states
between 400 and 800~MeV demonstrates the persistence
of the second resonance bump when at least one neutral meson is observed.

The current results indicate large differences between quasi\-free meson
production from the nuclear surface and non-quasifree components.
The quasifree part does not show a suppression of the resonance structures
in the second resonance region.
However, resonance structures seem absent in the
non-quasifree meson production which has larger contributions from the
nuclear volume.

It has to be concluded that the medium modifications leading to the depletion of
cross section in nuclear photoabsorption are a subtle interplay of effects.
Their investigation and the rigorous comparison to theoretical models requires
the detailed study of differential cross sections and a deeper understanding of
meson production in the nuclear medium.


\end{document}